\documentclass[fleqn,12pt,twoside]{article}
\usepackage{espcrc1}

\usepackage{graphicx}
\usepackage[figuresright]{rotating}

\newcommand{\AmS}{{\protect\the\textfont2
  A\kern-.1667em\lower.5ex\hbox{M}\kern-.125emS}}

\title{Supernova Neutrino Process and its Impact on the Galactic
Chemical Evolution of the Light Elements}

\author{
Takashi Yoshida\address{Astronomical Institute, Graduate School of Science, 
Tohoku University, \\
Aramaki, Aoba-ku, Sendai 980-8578, Japan},
Toshitaka Kajino\address{National Astronomical Observatory of Japan 
and The Graduate University for Advanced Studies, 
2-21-1 Osawa, Mitaka, Tokyo 181-8588, Japan}\address{Department of Astronomy, 
Graduate School of Science, University of Tokyo, \\
7-3-1 Hongo, Bunkyo-ku, Tokyo 113-0033, Japan}}
       
\begin{document}

\maketitle

\begin{abstract}
In order to resolve the overproduction problem of $^{11}$B in supernova
explosions during Galactic chemical evolution, the dependence
of the ejected masses of the light elements produced through the 
$\nu$-process in supernova explosions on supernova neutrino parameters
is investigated and constraints on the supernova neutrinos are evaluated.
Detailed nucleosynthesis in a supernova explosion model corresponding to 
SN 1987A is calculated by postprocessing.
The ejected masses of $^{11}$B and $^7$Li depend strongly on the temperature
of $\nu_{\mu,\tau}$ and $\bar{\nu}_{\mu,\tau}$ and are roughly proportional 
to the total neutrino energy.
The range of temperature of $\nu_{\mu,\tau}$ and $\bar{\nu}_{\mu,\tau}$
appropriate for the amount of $^{11}$B necessary for Galactic chemical 
evolution and the total neutrino energy deduced from the gravitational energy
of a typical neutron star is between 4.8 MeV and 6.6 MeV.
In the case of neutrino energy spectra with non-zero chemical potential,
this range decreases by about 10 \%.
\end{abstract}

\section{INTRODUCTION}

Supernova explosions are one of the sites promoted for the production of 
light elements, Li, Be, and B, in Galactic chemical evolution (GCE),
in addition to the Galactic Cosmic Rays (GCRs), AGB stars,and novae 
\cite{fo00}.
They continuously provide $^7$Li and $^{11}$B through the interactions
of nuclei such as $^4$He and $^{12}$C with neutrinos emitted from a 
proto-neutron star.
This synthesis process is called the $\nu$-process \cite{wh90}.

Overproduction of $^{11}$B in supernovae is one of the
standing problems in supernova nucleosynthesis and the GCE (e.g., \cite{fo00}).
Previous works have indicated that the supernova contribution of $^{11}$B 
production evaluated from supernova nucleosynthesis models \cite{ww95} is 
too large by a factor of 2.5 $\sim$ 5.6 to reproduce the GCE of the light 
elements; the factor depends on models of the GCE \cite{fo00,rl00,al02}.
Most supernova nucleosynthesis calculations were carried out with 
a total neutrino energy of $3 \times 10^{53}$ ergs and temperature of 8 MeV 
for $\nu_{\mu,\tau}$ and $\bar{\nu}_{\mu,\tau}$ and 4 MeV for $\nu_{\rm e}$ 
and $\bar{\nu}_{\rm e}$ \cite{wh90,ww95}.
The supernova explosion mechanism has not been resolved, so 
the characteristics of the supernova neutrinos, especially their energy
spectra, have not been fully determined \cite{kr03}.
Therefore, the light element amounts have not been obtained from a specific 
neutrino model precisely determined by supernova explosion models. 
Therefore, we systematically investigate the dependence of the light 
element synthesis in supernova explosions on the supernova neutrinos.
Furthermore, on the basis of the evaluated dependence and the results of the 
GCE of the light elements, we restrict the characteristics of the
supernova neutrinos, such as their energy spectra.


\section{SUPERNOVA EXPLOSION AND SUPERNOVA NEUTRINO MODELS}

The neutrino luminosity is assumed to decrease exponentially with a decay
time of $\tau_\nu = 3$ s as in \cite{wh90,ww95,yt04}.
The luminosity is equally divided into each favor of neutrinos.
The neutrino energy spectra are assumed to obey Fermi distribution with
zero chemical potential.
We set the temperature of $\nu_{\mu,\tau}$ and $\bar{\nu}_{\mu,\tau}$,
$T_\nu$, and the total neutrino energy $E_\nu$ as parameters.
The ranges of $T_\nu$ and $E_\nu$ are 4.0 MeV $\le T_\nu \le$ 9.0 MeV and
$1.0 \times 10^{53}$ ergs $\le E_\nu \le 6.0 \times 10^{53}$ ergs.
These ranges include the neutrino temperatures adopted in previous studies 
\cite{ww95,yt04} and the total neutrino energy range 
($2.4 \times 10^{53}$ ergs $\le E_\nu \le 3.5 \times 10^{53}$ ergs) 
deduced from the gravitational energy of a $\sim$ 1.4 $M_\odot$ neutron star 
\cite{lp01}.
For the temperatures of $\nu_{\rm e}$ and $\bar{\nu}_{\rm e}$, we choose
3.2 MeV and 5.0 MeV \cite{yt04}.
With these neutrino temperatures, a successful $r$-process abundance pattern,
an appropriate third-to-second peak ratio, 
has been obtained using a neutrino-driven wind model \cite{yt04}.

The supernova explosion is pursued using a spherically symmetrical 
Lagrangian PPM code with 13 element $\alpha$-particle nuclear reaction network 
for energy generation \cite{sn92}.
The presupernova structure is adopted from a 16.2 $M_\odot$ star corresponding
to SN 1987A \cite{sn90}.
The explosion energy is set to be $1 \times 10^{51}$ ergs and the mass cut
is set to be 1.6 $M_\odot$.
Detailed nucleosynthesis of the supernova explosion is calculated by
postprocessing.
The nuclear reaction network consists of 291 species of nuclei \cite{yt04}.
We interpolate the logarithmic values of the cross sections listed in 
\cite{hw92}.

\section{RESULTS AND DISCUSSION}

\begin{figure}[htb]
\begin{minipage}[t]{76mm}
\includegraphics[width=76mm]{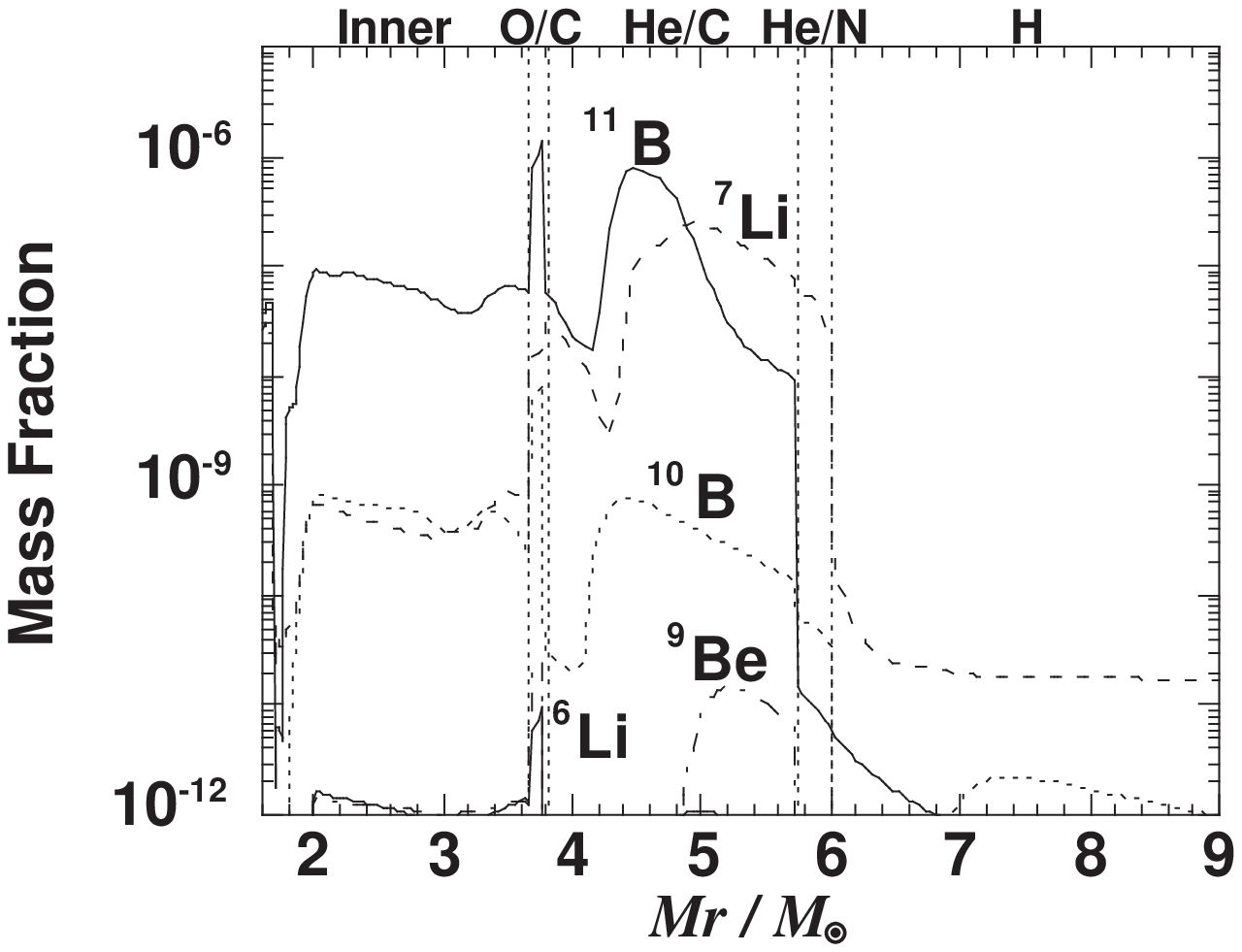}
\vspace*{-8mm}
\caption{
Mass fraction distribution of light elements (Li, Be, and B) in the case
of $E_\nu = 3 \times 10^{53}$ ergs and $T_\nu = 6$ MeV.
The mass fraction of $^7$Li is the sum of the mass fractions of $^7$Li and
$^7$Be.
The mass fraction of $^{11}$B is the sum of the mass fractions of $^{11}$B
and $^{11}$C.
}
\label{fig:f1}
\end{minipage}
\hspace{\fill}
\begin{minipage}[t]{76mm}
\includegraphics[width=76mm]{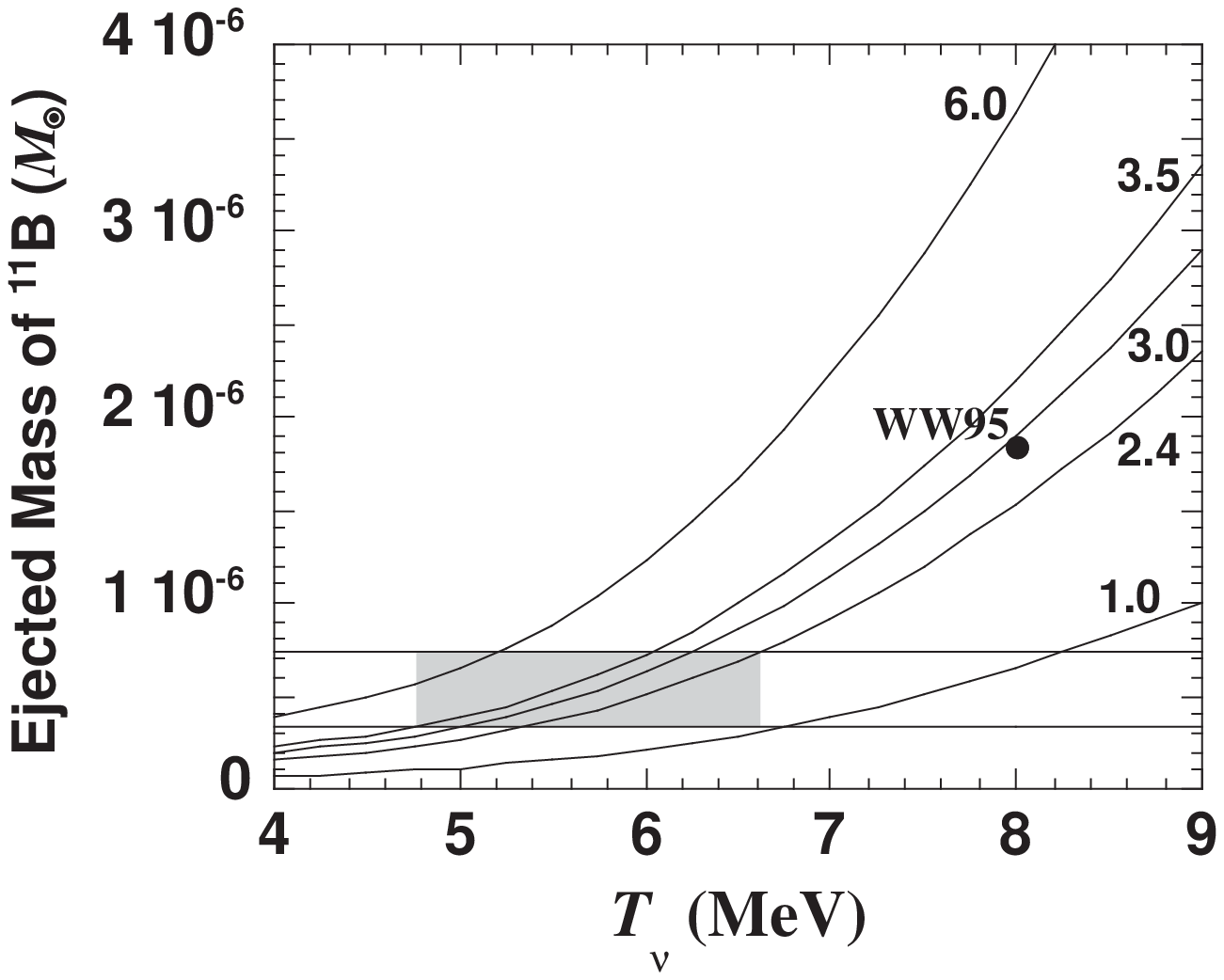}
\vspace*{-8mm}
\caption{
The ejected mass of $^{11}$B as a function of $T_\nu$.
Each attached number denotes $E_\nu$ in units of $1 \times 10^{53}$ ergs.
Two horizontal lines indicate the proper $^{11}$B mass range evaluated from 
the GCE models \cite{fo00,rl00,al02}.
The shaded region shows $T_\nu$ satisfying the proper ranges of the $^{11}$B
mass and $E_\nu$.
A point denoted by WW95 is the $^{11}$B ejected mass of S20A 
model \cite{ww95}. 
}
\end{minipage}
\end{figure}

Figure 1 shows mass fraction distribution of light elements 
in the supernova ejecta.
The light elements are mainly produced in the He/C layer 
and smaller amounts are in the inner O rich layer.
We have shown that $^7$Li and $^{11}$B are produced through the $\nu$-process
reactions: $^4$He($\nu,\nu'p)^3$H, $^4$He($\nu,\nu'n)^3$He and 
$\alpha$-captures in the He layer and $^{12}$C($\nu,\nu'p)^{11}$B in the 
inner O rich layer \cite{yt04}.
They are partly destroyed by the capture of $\alpha$-particles.

The ejected mass of $^{11}$B as a function of the
temperature of $\nu_{\mu,\tau}$ and $\bar{\nu}_{\mu,\tau}$, $T_\nu$, is 
shown in Fig. 2.
The dependence of $T_\nu$ on the ejected mass is stronger than linear. 
Since the ejected mass is roughly proportional to $E_\nu$ \cite{yt04}, 
the dependence of $T_\nu$ is stronger than the $E_\nu$ dependence.
This is because the dependence on the cross sections of the $\nu$-process
is larger than a linear dependence.
The $^{11}$B ejected mass in the case of $E_\nu=3 \times 10^{53}$ ergs and 
$T_\nu$ = 8 MeV is consistent with that in \cite{ww95} in spite of 
different values of $T_{\nu_{\rm e}}$ and $T_{\bar{\nu}_{\rm e}}$.
Since $T_{\nu_{\rm e}}$ and $T_{\bar{\nu}_{\rm e}}$ are smaller than 
$T_\nu$ and neutral current interactions are important for the
$\nu$-process of the light elements, the difference of $T_{\nu_{\rm e}}$
and $T_{\bar{\nu}_{\rm e}}$ from \cite{ww95} scarcely affects the ejected
masses of the light elements \cite{yt04}. 

We will apply the above results to GCE of the light elements.
Recent studies of the GCE have indicated that both GCRs and supernovae
contribute to the B production: $^{10}$B is produced through the
GCRs and $^{11}$B is produced through the GCRs and SNe.
In order to reproduce meteoritic $^{11}$B/$^{10}$B ratio (=4.05) at
solar metallicity, the supernova contribution of $^{11}$B is important since
the $^{11}$B/$^{10}$B ratio of the GCRs is 2.5 \cite{fo00,wm85}.
However, $^{11}$B is overproduced in SN nucleosynthesis models \cite{ww95}
compared to the evaluation of the GCE.
Prior evaluations of the GCE of the light elements give the overproduction 
factor $f_\nu$ to be 0.18 \cite{rl00}, 0.28 \cite{al02}, and 0.40 \cite{fo00}.
Thus, we set the range of the factor appropriate for the $^{11}$B amount 
in GCE to be $0.18 \le f_\nu \le 0.40$ as in \cite{yt04}.
The corresponding range of the $^{11}$B ejected mass is 
$3.3 \times 10^{-7} M_\odot \le M({\rm ^{11}B}) \le 
7.4 \times 10^{-7} M_\odot$ (two horizontal lines of Fig. 2).
We also restrict $E_\nu$ to be the gravitational energy of a typical neutron 
star (see \S 2) \cite{lp01}.
Finally, we obtain a shaded region satisfying both of the conditions
of the $^{11}$B mass and $E_\nu$ (see Fig. 2).
The obtained range of $T_\nu$ is 
$4.8 \, {\rm MeV} \, \le \, T_\nu \, \le \, 6.6 \, {\rm MeV}$, 
which is smaller than 8 MeV in \cite{ww95}.

Our evaluation is carried out only in the case of SN 1987A, 
which corresponds to about 20 $M_\odot$ star in ZAMS stage.
For $^{11}$B production, the $\nu$-process reactions of $^4$He in the He
layer and $^{12}$C in the O layer are important.
It depends on the progenitor mass models which layer is more important.
Fortunately, however, the neutrino temperature dependence does not change
strongly in either $\nu$-process reactions.
Thus, the appropriate range of the neutrino temperature would be a good 
approximation for supernovae with different progenitor masses.
The investigation of the dependence of the $^{11}$B production on the
neutrinos spectra for different progenitor mass is a future subject.

We briefly discuss the effect of non-zero chemical potential of the supernova
neutrinos.
Detailed discussion is written in \cite{yk04}.
Recent studies on the neutrino transfer during supernova explosions have
indicated that the energy spectra of the supernova neutrinos approximately
obey ^^ ^^ pinched'' Fermi-Dirac distribution rather than that with 
zero chemical potential \cite{kr03}.
In such a case, it is expected that the ejected masses of the light elements 
change since reaction rates of the $\nu$-process change.
Hartmann et al. \cite{hm99} discussed the effect of non-zero chemical 
potential assuming that cross sections of the $\nu$-process are proportional
to the square of the neutrino energy.
We generalize their discussion: we approximate the cross sections of the 
$\nu$-process to the $\alpha$th power low of the neutrino energy.
Then, we evaluate the effect on the reaction rates of the $\nu$-process.
The reaction rates of the $\nu$-process can be written as a function of the
degeneration factor $\eta_\nu=\mu_\nu/kT_\nu$ for given $\alpha$ and the
neutrino temperature $T_\nu$, and increase by a factor of 
1.4 $\sim$ 1.5 in the case of $\eta_\nu=3$ and of the $\alpha$ range 
$4 \le \alpha \le 7$ compared to that with $\eta_\nu=0$.
Since the ejected masses of $^{11}$B and $^7$Li are roughly proportional
to the $\nu$-process reaction rates, they increase with the same factor.
In this effect, the neutrino temperature range appropriate for the $^{11}$B 
amount in GCE should decrease in the case of non-zero chemical potential.
In the case of $\eta_\nu=3$, this range decreases by 10\% at most.

\end{document}